\documentclass[article,nojss]{jss}

\usepackage{amsmath}
\usepackage{amssymb}
\usepackage{float}

\newcommand{\file}[1]{\texttt{#1}}

\author{
  Austin Talbot\\Emory University
  \And
  Ilha Hwang\\Brigham Young University
  \And
  Cristina Trevino\\Emory University
  \And
  Alex Kotlar\\Emory University
}

\title{\pkg{scikit-covtest}: Covariance Matrix Hypothesis Testing in Python}

\Plainauthor{Austin Talbot, Ilha Hwang, Cristina Trevino, Alex Kotlar}
\Plaintitle{scikit-covtest: Covariance Matrix Hypothesis Testing in Python}
\Shorttitle{\pkg{scikit-covtest}: Covariance Matrix Hypothesis Testing}

\Abstract{
Covariance matrices play a central role across diverse domains such as finance,  machine learning, neuroscience, and genetics, where they are used for tasks  including dimensionality reduction, connectivity inference, and risk estimation.  Many of these applications require testing whether a covariance matrix follows  a specific structure. While several \proglang{R} packages provide partial coverage of such  tests, \proglang{Python}, an important language in machine learning, lacks a comprehensive,  well-tested implementation. To address this gap, we introduce \pkg{scikit-covtest},  a Python package implementing a variety of hypothesis tests for covariance matrices  spanning four categories: identity, sphericity, proportionality, and two-sample  equality. The package provides a consistent \pkg{scipy}-style API, extensive  documentation, and supporting functionality for multiple testing correction,  synthetic data generation, and diagnostic evaluation. \pkg{scikit-covtest} is open source, available through PyPI, and lowers the
barrier to applying modern covariance-testing methods in scientific
applications.

}


\Plainkeywords{
  covariance matrix,
  hypothesis testing,
  high-dimensional statistics,
  multiple hypothesis testing,
  Python}

\Address{
  Austin Talbot\\
  Department of Human Genetics\\
  Emory University\\
  Atlanta, Georgia\\
  E-mail: \email{atalbo4@emory.edu}\\[1ex]
  Ilha Hwang\\
  Brigham Young University\\
  Provo, Utah\\[1ex]
  Cristina Trevino\\
  Department of Human Genetics\\
  Emory University\\
  Atlanta, Georgia\\
  E-mail: \email{ctrevin@emory.edu}\\[1ex]
  Alex Kotlar\\
  Department of Human Genetics\\
  Emory University\\
  Atlanta, Georgia\\
  E-mail: \email{akotlar@emory.edu}
}

\begin{document}

\section[Introduction]{Introduction}
\label{sec:introduction}

Covariance matrices play a central role in modern statistics, with applications spanning finance~\citep{pafka2003noisy}, machine learning~\citep{minh2017covariances}, neuroscience~\citep{varoquaux2010brain}, and genetics~\citep{head2023poirot,hwang2026poise}. They are used for dimensionality reduction~\citep{bishop2006pattern}, network discovery~\citep{zhong2009detecting}, and quantifying risk~\citep{ledoit2022power}. In genetics, covariance matrices are critical in characterizing the relationships between genetic variants~\citep{xu2013principles}. As one of the fundamental quantities in the field, they are used in a variety of tasks such as gene-trait associations~\citep{li2023mbat}, polygenic risk score estimation~\citep{ge2019polygenic}, and fine mapping~\citep{wang2020simple}. Hypothesis testing of covariance matrix structure is therefore a fundamental task in these domains, both in the classical large sample size domain and the high-dimensional domain where the number of variables is comparable or exceeds sample size.

Despite the importance of testing, and the extensive work developing methods for doing so, software implementations of these methods are fragmented and incomplete. In \proglang{R}, several packages provide implementations of specific procedures~\citep{ding2025package,cao2018package,barnard2018covtestr}. However, these packages omit many methods, maintenance levels vary, and some have unresolved implementation issues documented in their repositories. In \proglang{Python}, functionality is even more limited: \pkg{statsmodels}~\citep{seabold2010statsmodels} includes a small number of likelihood-based procedures and \pkg{Pingouin}~\citep{vallat2018pingouin} implements the classic Box's M test~\citep{box1953non}, but to our knowledge, there is no unified, well-tested, and well-documented framework dedicated to covariance matrix hypothesis testing.

We introduce \pkg{scikit-covtest}, a \proglang{Python} package that provides a comprehensive, consistent, and tested implementation of a broad class of tests. These tests are organized into four main categories covering most common testing scenarios: (i) one-sample tests of a specified covariance matrix, (ii) sphericity tests, (iii) proportionality tests between two covariance matrices, and (iv) two-sample tests for equality of covariance matrices. The design follows a \pkg{SciPy}-style API, with standardized input conventions, clear return types, extensive documentation, and rigorous testing. 

Beyond hypothesis tests, \pkg{scikit-covtest} includes supporting functionality essential in many applications. Multiple testing corrections~\citep{benjamini1995controlling,storey2003statistical,benjamini2001control} are critical when many tests are conducted simultaneously, a common occurrence in the era of genome-wide studies. Synthetic data generation with controlled covariance structures enables reproducible benchmarking and method validation. Diagnostic functions for assessing multivariate normality~\citep{mardia1970measures} are also included, as normality is a common and critical assumption in many implemented tests. The package also includes curated datasets~\citep{lecun2002gradient,fiorini2016gene}, benchmarking utilities to evaluate test properties, and visualization tools to facilitate empirical evaluation and interpretation of the results.

Due to the broad usage of covariance matrices, different tests have been designed for specific regimes and alternatives. The most important distinctions are (i) whether the test is either classical or high-dimensional and (ii) whether the test is tuned for sparse or dense alternatives. Tests designed for one regime may be poorly calibrated, undefined, or low-power in another. Furthermore, many classical tests assume multivariate normality but differ in sensitivity to violations of this assumption. To illustrate these differences, we provide the outputs of the tests under standardized simulations to assist in determining which tests are appropriate for a given application.

The contents of this paper are as follows. In Section~\ref{sec:related_work} we provide the mathematical background on hypothesis testing and describe the four classes of tests in this work. Section~\ref{sec:methods} describes the methods currently implemented in \pkg{scikit-covtest}, including the supporting methods for data generation, evaluation, and multiple hypothesis testing mentioned above. Section~\ref{sec:simulations} evaluates null calibration and empirical power across all four hypothesis classes. Section~\ref{sec:applied} demonstrates usage on two of the bundled datasets. Finally, Section~\ref{sec:conclusion} provides some suggestions on future improvements and work. This package is installable via pip and is publicly available at \url{https://github.com/bystrogenomics/scikit-covtest}, along with the code to reproduce all figures and analyses in this work.


\section[Background: Hypothesis Testing of Covariance Matrices]
{Background: Hypothesis Testing of Covariance Matrices}
\label{sec:related_work}

\subsection{Notation and Problem Formulation}
\label{ssec:2p1}

Vectors are denoted by bold lowercase letters (e.g., $\mathbf{x}$) and matrices 
by bold uppercase letters (e.g., $\mathbf{X}$). For a matrix $\mathbf{A} = (a_{ij})$, 
its transpose is $\mathbf{A}^\top$. The covariance matrix is denoted by 
$\boldsymbol{\Sigma} = (\sigma_{ij})$.

In the one-sample setting, let $\mathbf{x}_1, \dots, \mathbf{x}_n$ be i.i.d.\ 
$p$-dimensional random vectors with mean $\boldsymbol{\mu} = \mathbb{E}[\mathbf{x}_k]$ 
and covariance matrix $\boldsymbol{\Sigma} = \mathrm{Cov}(\mathbf{x}_k)$. The sample 
mean and covariance matrix are
\begin{equation}
\bar{\mathbf{x}} = \frac{1}{n}\sum_{k=1}^n \mathbf{x}_k, 
\qquad 
\widehat{\boldsymbol{\Sigma}} = \frac{1}{n}\sum_{k=1}^n (\mathbf{x}_k - \bar{\mathbf{x}})(\mathbf{x}_k - \bar{\mathbf{x}})^\top.
\end{equation}

In the two-sample setting, we observe independent samples 
$\{\mathbf{x}_1, \dots, \mathbf{x}_{n_1}\}$ and $\{\mathbf{y}_1, \dots, \mathbf{y}_{n_2}\}$ 
with covariance matrices $\boldsymbol{\Sigma}_1$ and $\boldsymbol{\Sigma}_2$, and 
corresponding estimators $\widehat{\boldsymbol{\Sigma}}_1$ and $\widehat{\boldsymbol{\Sigma}}_2$.

We consider the following hypotheses for testing covariance structures:
\begin{itemize}
    \item \textbf{Identity test:} 
    Tests whether the population covariance matrix equals the identity,
    \begin{equation}
        H_0:\ \boldsymbol{\Sigma} = \mathbf{I}_p,
    \end{equation}
    which corresponds to variables being uncorrelated with unit variances; note that 
    this constrains only the second moments and does not restrict the mean. This 
    generalizes to $H_0: \boldsymbol{\Sigma} = \boldsymbol{\Sigma}_0$ for known 
    $\boldsymbol{\Sigma}_0$ either by replacing $\mathbf{I}_p$ with 
    $\boldsymbol{\Sigma}_0$ or, equivalently, by whitening the data through 
    $\boldsymbol{\Sigma}_0^{-1/2}\widehat{\boldsymbol{\Sigma}}\boldsymbol{\Sigma}_0^{-1/2}$ 
    and testing against $\mathbf{I}_p$, with calibration following the transformed 
    statistic. 
    Applications include monitoring for changes in machine 
    alignment~\citep{li2013monitoring} and testing for intrasubject variation in 
    high-dimensional gene expression data~\citep{qayed2021high}.

    \item \textbf{Sphericity test:} 
    Tests whether the covariance matrix is proportional to the identity,
    \begin{equation}
        H_0:\ \boldsymbol{\Sigma} = c\,\mathbf{I}_p,\ \ c>0,
    \end{equation}
    implying that all variables have equal variance and are mutually uncorrelated. 
    Sphericity tests arise in repeated-measures ANOVA, where $c\,\mathbf{I}_p$ is a 
    sufficient condition for valid within-subject comparisons (the exact requirement 
    being sphericity of the within-subject contrasts)~\citep{winer1971statistical,smith2017guerilla}. 
    They also appear in signal processing applications such as determining the 
    number of signal sources in radar systems~\citep{liu2017source,yuan2023rmt}.

    \item \textbf{Proportionality test:} 
    Tests whether two covariance matrices differ only by a scalar multiple,
    \begin{equation}
        H_0:\ \boldsymbol{\Sigma}_1 = c\,\boldsymbol{\Sigma}_2,\ \ c>0,
    \end{equation}
    indicating the same correlation structure but potentially different overall scale. 
    This has been used to test genetic variance matrices as an intermediate hypothesis 
    between complete equality and completely unrelated 
    structures~\citep{phillips1999hierarchical} and to test specific evolutionary 
    patterns when overall variance is not of interest~\citep{ackermann2002patterns}.

    \item \textbf{Two-sample equality test:} 
    Tests whether two populations share the same covariance structure,
    \begin{equation}
        H_0:\ \boldsymbol{\Sigma}_1 = \boldsymbol{\Sigma}_2,
    \end{equation}
    corresponding to equal variance and correlation patterns across samples. 
    Applications include detecting genotype-phenotype 
    interactions~\citep{wang2019genotype}, identifying parent-of-origin 
    effects~\citep{head2023poirot}, and detecting changes in financial portfolio 
    variances~\citep{li2025tests}.
\end{itemize}

\subsection{Common Test Statistics}

Many modern hypothesis tests are based on norms to measure distance between 
the sample covariance(s) and the hypothesized structure. For example, a single-sample 
test would be
\begin{equation}\label{eq:single_sample}
T=\bigl\|\widehat{\boldsymbol{\Sigma}}-\mathbf{I}_p\bigr\|^2.
\end{equation}
The choice of norm determines the alternatives that the test is sensitive to. For 
example, the Frobenius norm is sensitive to dense alternatives where a large 
proportion of entries differ from the null. This behavior can be seen when this 
norm is written in its alternate form as
\begin{equation}
\bigl\|\widehat{\boldsymbol{\Sigma}}-\mathbf{I}_p\bigr\|_F^2
=\sum_{i=1}^p (\widehat{\sigma}_{ii}-1)^2
+2\sum_{1\le i<j\le p}\widehat{\sigma}_{ij}^2.
\end{equation}
Alternatively, the operator norm,
\begin{equation}
\bigl\|\widehat{\boldsymbol{\Sigma}}-\mathbf{I}_p\bigr\|_{\mathrm{op}}
=\max_{\|\mathbf{u}\|_2=1}\,\bigl|\mathbf{u}^\top(\widehat{\boldsymbol{\Sigma}}-\mathbf{I}_p)\mathbf{u}\bigr|,
\end{equation}
will target departures in a single direction and is sensitive to low-rank structure. 
Finally, the entrywise maximum norm $\|\mathbf{A}\|_{\max}=\max_{1\le i,j\le p}|a_{ij}|$ 
(the $L_\infty$ norm of the vectorized matrix) targets sparse deviations. Two-sample 
tests function similarly, except the norm is taken between the two empirical covariance 
matrices. 

\begin{figure}[t!]
\centering
\includegraphics[width=\textwidth]{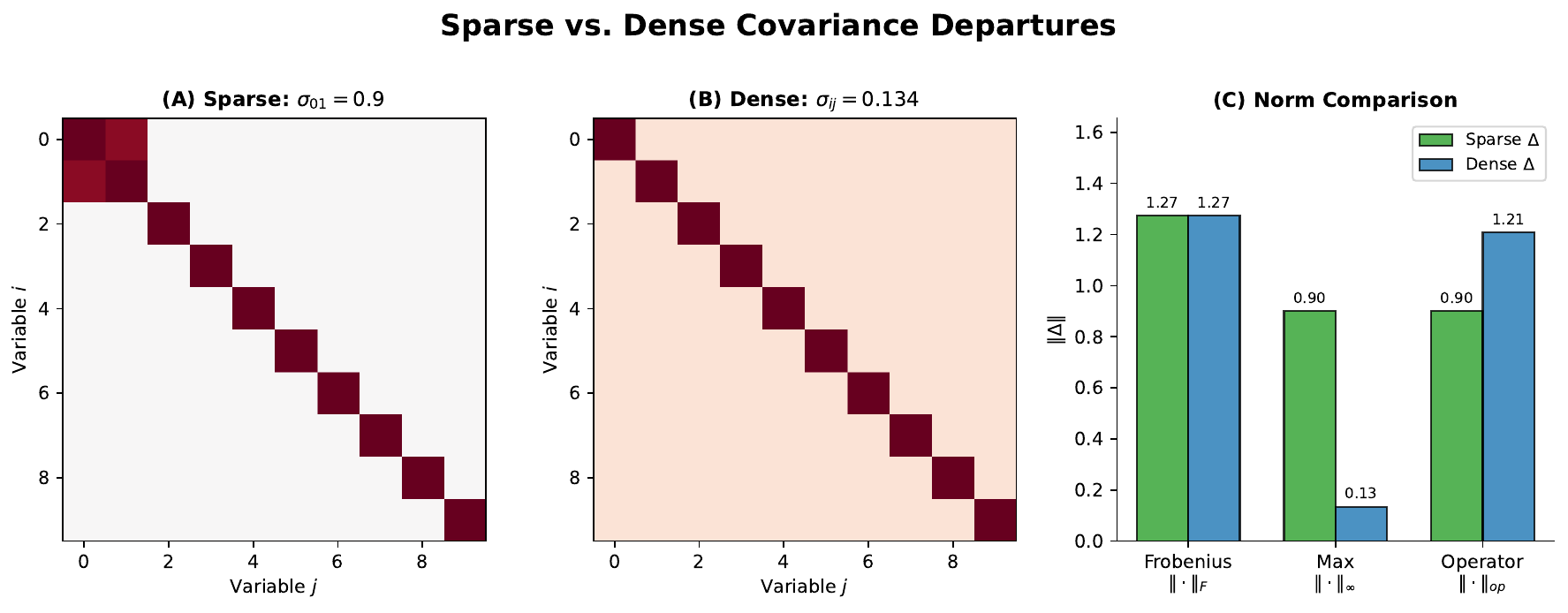}
\caption{\label{fig:norms} Illustration of sparse versus dense alternatives to the 
identity hypothesis, shown for $p=10$ with departure matrix 
$\boldsymbol{\Delta}=\boldsymbol{\Sigma}-\mathbf{I}_p$. (A) Sparse alternative with a 
single large off-diagonal correlation ($\sigma_{12}=0.9$). (B) Dense alternative with 
uniform small correlations ($\sigma_{ij}=0.134,\;i\ne j$), calibrated to have the same 
Frobenius norm as (A). (C) Comparison of matrix norms. The two departure matrices have 
equal Frobenius norm by construction, but the entrywise maximum norm is $6.7\times$ 
larger for the sparse case, while the operator norm is $1.3\times$ larger for the dense 
case. This illustrates why entrywise-maximum ($L_\infty$) based tests are preferred for 
detecting sparse departures, while Frobenius-based tests are sensitive to dense 
alternatives and operator-norm tests focus on low-rank structure.}
\end{figure}

Sphericity tests require modification to account for the unknown scale $c$. Centering 
the sample covariance about its best scalar approximant removes the identity component 
but not the scale itself, so the statistic must additionally be normalized by an 
estimate of $c$ to obtain a null distribution free of the nuisance scale. Define the 
centered sample covariance matrix
\begin{equation}
\widetilde{\boldsymbol{\Sigma}}
=
\widehat{\boldsymbol{\Sigma}}
-
\frac{\operatorname{tr}\!\left(\widehat{\boldsymbol{\Sigma}}\right)}{p}
\mathbf{I}_p,
\qquad
\widehat{c}=\frac{\operatorname{tr}\!\left(\widehat{\boldsymbol{\Sigma}}\right)}{p}.
\end{equation}
A scale-invariant Frobenius-type sphericity statistic is
\begin{equation}
T_F^{\mathrm{sph}}
=
\frac{\bigl\|\widetilde{\boldsymbol{\Sigma}}\bigr\|_F^2}{p\,\widehat{c}^{\,2}}
=
\frac{p\operatorname{tr}\!\left(\widehat{\boldsymbol{\Sigma}}^2\right)}
{\operatorname{tr}\!\left(\widehat{\boldsymbol{\Sigma}}\right)^2}-1,
\end{equation}
which coincides with the classical statistics of \citet{john1971some} and 
\citet{ledoit2002some} and whose null distribution does not depend on $c$. 
Max-type sphericity tests target sparse alternatives by examining the largest 
standardized entry of $\widetilde{\boldsymbol{\Sigma}}$,
\begin{equation}
T_{\max}^{\mathrm{sph}}
=
\max_{1\le i\le j\le p}
\frac{\bigl|\widetilde{\sigma}_{ij}\bigr|}
{\widehat{\operatorname{se}}\!\left(\widetilde{\sigma}_{ij}\right)},
\end{equation}
where each entry is divided by an estimate of its null standard error. This 
standardization is necessary because, under sphericity, diagonal entries of the 
sample covariance have approximately twice the variance of the off-diagonal entries, 
so an unstandardized maximum would combine non-comparable coordinates; the 
standardized maximum is scale-invariant and is calibrated using an extreme-value 
(Gumbel) limit~\citep{cai2013two}. Figure~\ref{fig:norms} illustrates the distinction 
between norms: two alternatives with identical Frobenius norms can have substantially 
different maximum-entry and operator norms.

In the classical regime where $n$ is large relative to $p$, traditional distributional 
convergence, either to a chi-squared~\citep{ledoit2002some,hallin2006semiparametrically,muirhead2009aspects} 
or normal~\citep{ahmad2015tests,li2025tests} distribution, can be used. However, 
when the data are high dimensional, tests often require bias correction on the test statistic to obtain 
proper size control and power~\citep{ledoit2002some,srivastava2007multivariate}.

Beyond distributional assumptions, test robustness to heavy tails is often desirable. 
Robust variants replace $\widehat{\boldsymbol{\Sigma}}$ by heavy-tail-resistant 
scatter matrices, including procedures inspired by robust $T^2$~\citep{o1992robust} 
and tests based on Tyler's $M$-estimator~\citep{li2025tests}. Because Tyler's estimator 
identifies the elliptical shape only up to a positive scalar, such procedures are 
naturally suited to normalized shape, sphericity, and proportionality hypotheses; 
testing an absolute-scale null additionally requires a separate robust scale estimate.

\subsection{Hypothesis Testing Diagnostics}

To evaluate test calibration, we implement diagnostics for p-value uniformity under 
the null hypothesis. Classical goodness-of-fit tests assess whether p-values behave 
as expected under the null. The Kolmogorov-Smirnov test compares the empirical 
distribution to $\mathrm{Uniform}(0,1)$, while an Anderson-Darling 
statistic~\citep{lehmann2005testing} targets deviations in the distribution tails. 
Scalar measures quantify global discrepancies between the empirical and theoretical 
cumulative distribution functions. We note that exact uniformity is expected only for 
continuous, exactly calibrated p-values; discrete, permutation, and Monte Carlo 
p-values are typically only super-uniform, and the standard reference distributions 
for these goodness-of-fit statistics assume independence across the tested hypotheses.

Large-scale testing applications require additional diagnostics. The genomic control 
inflation factor $\lambda_{\mathrm{GC}}$~\citep{devlin1999genomic} detects systematic 
test statistic inflation via the median $\chi^2$ statistic. Storey's $\pi_0$ 
estimator~\citep{storey2003statistical} infers the proportion of true nulls by 
examining the tail behavior of p-values. Both are descriptive summaries whose 
interpretation depends on signal prevalence: inflation or a low estimated $\pi_0$ can 
reflect widespread true signal rather than miscalibration alone. Together, these 
methods provide both formal tests and descriptive summaries for evaluating test 
behavior in high-dimensional settings.

\subsection{Multiple Hypothesis Testing}

Multiple hypothesis testing is essential in high-dimensional data analysis, 
particularly in omics fields where researchers routinely evaluate thousands to 
millions of hypotheses~\citep{dudoit2008multiple}. In genome-wide association 
studies (GWAS), for example, millions of SNPs are tested simultaneously, and 
without appropriate corrections, false positives would overwhelm true 
findings~\citep{visscher2012five}. 

Standard correction methods include the false discovery rate (FDR)~\citep{benjamini1995controlling} 
and family-wise error rate (FWER) controls such as the Bonferroni adjustment. 
The widely used GWAS threshold of $p < 5 \times 10^{-8}$ exemplifies how 
multiple-testing corrections have been scaled to genomic 
contexts~\citep{pe2008estimation}. These procedures help ensure that high-throughput 
studies yield reproducible, reliable results, minimizing wasted resources on spurious 
associations~\citep{dudbridge2008estimation,storey2003statistical}.

To support both classical and modern testing contexts, the package implements 
several FDR control strategies, each valid under specific dependence conditions. The 
Benjamini-Hochberg (BH)~\citep{benjamini1995controlling} procedure controls the FDR 
under independence and under positive regression dependence on a subset (PRDS), while 
the Benjamini-Yekutieli (BY)~\citep{benjamini2001control} procedure guarantees control 
under arbitrary dependence at the cost of a logarithmic penalty factor. Extensions such 
as weighted BH~\citep{genovese2006false} incorporate prior information through 
nonnegative weights fixed independently of the p-values to improve power in 
heterogeneous settings. Storey-Tibshirani's adaptive method~\citep{storey2003statistical} 
further increases power by estimating the null proportion $\pi_0$, under corresponding 
assumptions on the estimator and dependence structure.

\subsection{Testing Distributional Assumptions}

Many classical covariance tests assume Gaussian or elliptical distributions and 
well-conditioned covariance matrices. To evaluate these assumptions, we implement 
several diagnostic methods. Eigenvalue spectrum analysis detects deviations from white noise or low-rank structure. 
The package computes the sample covariance spectrum with optional Marchenko-Pastur 
bounds~\citep{marvcenko1967distribution} as an isotropic, white-noise reference for 
comparison. Multivariate normality is 
assessed through Mardia's skewness and kurtosis tests~\citep{mardia1970measures}, 
the Henze-Zirkler test~\citep{henze1990class}, and Royston's 
procedure~\citep{royston1983some}. These tests combine marginal normality with 
inter-variable correlation structure to provide complementary assessments of 
multivariate Gaussianity; because they rely on a nonsingular sample scatter matrix, 
they apply in the classical regime $p<n$ and require regularization or dimension 
reduction when $p \ge n$. Matrix conditioning is evaluated via the spectral condition number, numerical rank, 
and entropy-based effective rank~\citep{roy2007effective}. Warnings are issued for 
ill-conditioned, rank-deficient, or strongly collinear matrices, which can compromise 
test performance and lead to unreliable inference.

\section[scikit-covtest]{scikit-covtest}
\label{sec:methods}

\begin{figure}
  \centering
  \includegraphics[width=\linewidth]{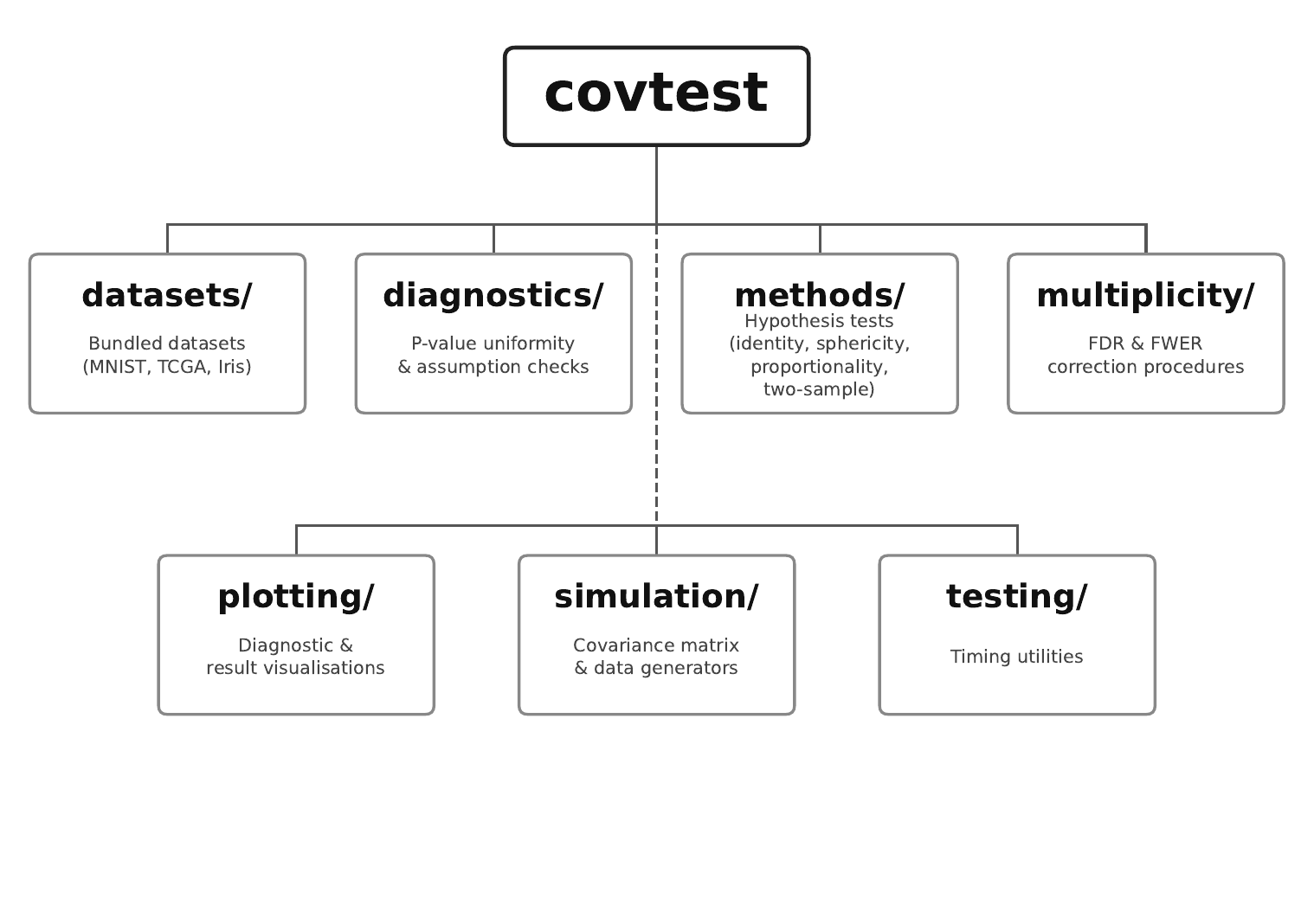}
\caption{Directory structure of the \pkg{covtest} package. The root package
contains seven submodules organised into two groups. The core group comprises
\code{datasets} (bundled benchmark datasets: MNIST, TCGA, and Iris),
\code{diagnostics} (p-value uniformity and distributional assumption checks),
\code{methods} (implementations of hypothesis tests for identity, sphericity,
proportionality, and two-sample covariance hypotheses), and
\code{multiplicity} (FDR and FWER correction procedures). The support group
comprises \code{plotting} (diagnostic and result visualisations),
\code{simulation} (covariance matrix and data generators), and \code{testing}
(internal timing utilities).}
  \label{fig:repo_structure}
\end{figure}

\subsection{Testing Methods}
scikit-covtest version 0.1.0 is available on PyPI (pip install scikit-covtest) and on GitHub at \url{https://github.com/bystrogenomics/scikit-covtest} under the MIT license. All examples in this paper were produced with this version.

All hypothesis tests in \pkg{scikit-covtest} follow a single, lightweight interface aligned with \pkg{SciPy} style. Each function accepts an $n \times p$ data matrix and optional arguments that specify the null structure, and returns a Python dict with two keys: \texttt{stat} and \texttt{p\_value}. For example, \code{ledoit_wolf_identity(X)} computes the test statistic under an identity null and returns \texttt{\{\,"stat": \dots, "p\_value": \dots\,\}}. Inputs may be NumPy arrays or pandas DataFrames, and outputs are plain Python types that serialize cleanly for reproducible workflows.

The shared return format enables seamless composition with familiar tooling. Results can be tabulated across parameter grids, plotted against sample size or dimension, and passed into multiple testing procedures using the same field names. Because every test returns \texttt{stat} and \texttt{p\_value}, benchmarking and automated reporting can be written once and reused across the library. This design lowers cognitive overhead, supports transparent evaluation, and encourages side-by-side comparisons of classical and high-dimensional procedures under a single, consistent API.

We have evaluated the methods implemented in \pkg{scikit-covtest}
using standardized simulation studies. A representative subset of
these evaluations is presented in Section~4, while complete,
reproducible notebooks are available in the GitHub repository.
Table~1 summarizes the currently implemented procedures across the
four hypothesis classes considered in this package: identity,
sphericity, proportionality, and two-sample covariance equality.
For each method, the table indicates whether it is applicable in
high-dimensional settings and the type of alternative to which it is
primarily sensitive. As additional methods are incorporated, the same
benchmarking framework can be used to extend these comparisons.


\begin{table}[H]
\centering
\small
\renewcommand{\arraystretch}{1.08}
\caption{Covariance structure tests by high-dimensional applicability
and alternative sensitivity}
\label{tab:tests_all}
\begin{tabular}{p{8.0cm} c c}
\hline
\textbf{Test}
& \textbf{High Dimensional}
& \textbf{Alternative Sensitivity} \\
\hline

\multicolumn{3}{l}{\textbf{Identity Tests}} \\
\cite{nagao1973some}
    & No
    & Dense \\
\cite{ledoit2002some}
    & Yes
    & Dense \\
\cite{srivastava2005some}
    & Yes
    & Dense \\
\cite{chen2010tests}
    & Yes
    & Dense \\
\cite{srivastava2011some}
    & Yes
    & Dense \\
\cite{fisher2012testing}
    & Yes
    & Dense \\
\cite{srivastava2014tests}
    & Yes
    & Dense \\
\cite{ahmad2015tests}
    & Yes
    & Dense \\
\cite{xu2025adjusted}
    & Yes
    & Dense \\

\multicolumn{3}{l}{\textbf{Sphericity Tests}} \\
\cite{bartlett1937properties}
    & No
    & Dense \\
\cite{john1971some}
    & No
    & Dense \\
\cite{srivastava2005some}
    & Yes
    & Dense \\
\cite{muirhead2009aspects}
    & No
    & Dense \\
\cite{fisher2010new}
    & Yes
    & Low-rank/spiked \\
\cite{chen2010tests}
    & Yes
    & Dense \\
\cite{srivastava2014tests}
    & Yes
    & Dense \\
\cite{ahmad2015tests}
    & Yes
    & Dense \\
\cite{xu2025adjusted}
    & Yes
    & Dense \\

\multicolumn{3}{l}{\textbf{Proportionality Tests}} \\
\cite{eriksen1987proportionality}
    & No
    & Dense \\
\cite{liu2014new}
    & Yes
    & Dense \\
\cite{cheng2019testing}
    & Yes
    & Dense \\
\cite{tsukuda2019high}
    & Yes
    & Dense \\
\cite{ahmad2022tests}
    & Yes
    & Dense \\

\multicolumn{3}{l}{\textbf{Two-Sample Tests}} \\
\cite{box1953non}
    & No
    & Dense \\
\cite{schott2001some}
    & No
    & Dense \\
\cite{schott2007test}
    & Yes
    & Dense \\
\cite{srivastava2007multivariate}
    & Yes
    & Dense \\
\cite{srivastava2014tests}
    & Yes
    & Dense \\
\cite{cai2013two}
    & Yes
    & Sparse \\
\cite{ishii2017high}
    & Yes
    & Low-rank/spiked \\
\cite{chang2017comparing}
    & Yes
    & Sparse \\
\cite{ahmad2017testing}
    & Yes
    & Dense \\
\cite{he2018high}
    & Yes
    & Sparse \\
\cite{ding2024two}
    & Yes
    & Low-rank/spiked \\
\cite{li2025tests}
    & Yes
    & Dense \\

\hline
\end{tabular}
\end{table}

\subsection{Multiple Hypothesis Testing}

Multiple-hypothesis testing is a critical component of modern
high-dimensional data analysis because it provides a framework for
distinguishing reproducible signals from noise when many hypotheses are
tested simultaneously. This issue arises naturally when covariance
tests are applied across many genes, feature blocks, experimental
conditions, or population comparisons. Without multiplicity correction,
the probability of reporting false-positive results increases rapidly
with the number of tests.

The problem is especially prominent in omics applications, where
researchers routinely perform thousands or millions of association
tests. In genome-wide association studies (GWAS), for example, millions
of single nucleotide polymorphisms may be evaluated for association
with a phenotype. The conventional genome-wide significance threshold
of $p < 5 \times 10^{-8}$ reflects the adaptation of family-wise
error-rate control to this testing scale
\citep{pe2008estimation}. More generally, appropriate control of the
false discovery rate (FDR) or family-wise error rate (FWER) helps ensure
that high-throughput studies produce reliable and reproducible findings
\citep{dudbridge2008estimation,storey2003statistical}.

\subsubsection{False Discovery Rate}

The FDR is the expected proportion of false rejections among all
rejected hypotheses. The Benjamini--Hochberg procedure
\citep{benjamini1995controlling} controls the FDR under independence and
certain positive-dependence conditions. The Benjamini--Yekutieli
procedure \citep{benjamini2001control} introduces a harmonic correction
that provides control under arbitrary dependence, although the
additional correction generally reduces power.

The module also implements procedures for incorporating external
information or estimating the proportion of true null hypotheses. The
weighted Benjamini--Hochberg procedure
\citep{genovese2006false} assigns nonnegative weights to the hypotheses,
allowing hypotheses with stronger prior support to receive more
favorable rejection thresholds. The implementation normalizes the
weights internally so that they sum to the number of hypotheses.

The Storey--Tibshirani procedure
\citep{storey2003statistical} estimates the proportion $\pi_0$ of true
null hypotheses from the upper tail of the observed $p$-value
distribution. It evaluates the estimate over a user-specified or
default grid of tuning parameters $\lambda$ and uses the resulting
estimate to calculate adjusted values. When $\widehat{\pi}_0 < 1$, this
can increase power relative to procedures that implicitly assume that
all hypotheses are null.

Additional implemented methods include the step-down procedure of
\citet{benjamini1999step} and the weighted step-up construction of
\citet{blanchard2008two}. The latter accepts optional prior weights
through the argument \texttt{pii}. These weights influence the rejection
thresholds but do not constitute an estimate of $\pi_0$.

\subsubsection{Family-Wise Error Rate}

The FWER is the probability of making at least one false rejection
within a family of hypotheses. The Bonferroni and Holm procedures
provide strong FWER control without requiring independence among the
tests. Hochberg and Hommel procedures can provide greater power under
stronger assumptions on the dependence structure.

The Romano--Wolf procedure \citep{romano2005exact} differs from the
other implemented methods because it explicitly uses the joint null
distribution of the test statistics. Rather than accepting marginal
$p$-values, it takes observed test statistics and a matrix of
bootstrap or permutation statistics generated under the joint null.
This resampling approach can account for dependence among the
hypotheses, provided that the resampling scheme adequately represents
their joint null distribution.

Table~\ref{tab:multipletesting} summarizes the implemented procedures,
whether they estimate $\pi_0$, and their principal assumptions. A
general review of multiple-testing methodology is provided by
\citet{austin2014multiple}.

\begin{table}[H]
\centering
\caption{Multiple-testing procedures implemented in
\pkg{scikit-covtest}.}
\label{tab:multipletesting}
\small
\begin{tabular}{l c l}
\hline
\textbf{Procedure} &
\textbf{Estimates $\boldsymbol{\pi_0}$} &
\textbf{Main requirement} \\
\hline

\multicolumn{3}{l}{\textit{Family-wise error rate procedures}} \\[2pt]

\cite{dunn1961multiple}
& No
& Arbitrary dependence \\

\cite{holm1979simple}
& No
& Arbitrary dependence \\

\cite{hochberg1988sharper}
& No
& Independence or positive dependence \\

\cite{hommel1988stagewise}
& No
& Independence or positive dependence \\

\cite{romano2005exact}
& No
& Valid joint null resamples \\

\hline

\multicolumn{3}{l}{\textit{False discovery rate procedures}} \\[2pt]

\cite{benjamini1995controlling}
& No
& Independence or PRDS \\

\cite{benjamini2001control}
& No
& Arbitrary dependence \\

\cite{genovese2006false}
& No
& Fixed weights; independence or PRDS \\

\cite{storey2003statistical}
& Yes
& Independence \\

\cite{benjamini1999step}
& No
& Independence \\

\cite{blanchard2008two}
& No
& Positive prior weights; arbitrary dependence \\

\hline
\end{tabular}
\end{table}

The procedures are divided between the \texttt{fdr} and \texttt{fwer}
submodules of \texttt{covtest.multiplicity}. The FDR procedures
generally accept a vector of $p$-values and the desired error level
\texttt{alpha}. Weighted Benjamini--Hochberg additionally accepts
\texttt{weights}, the Blanchard--Roquain procedure accepts optional
prior weights through \texttt{pii}, and the Storey--Tibshirani
procedure accepts an optional grid of tuning parameters through
\texttt{lambdas}.

The Bonferroni, Holm, Hochberg, and Hommel FWER procedures similarly
accept marginal $p$-values and \texttt{alpha}. Romano--Wolf instead
accepts the observed statistics \texttt{T\_obs}, a matrix of jointly
resampled statistics \texttt{T\_boot}, and an optional \texttt{side}
argument specifying the direction of the alternative.

All procedures return a dictionary containing \texttt{rejected},
\texttt{alpha}, and \texttt{method}. The FDR procedures return adjusted
values under the key \texttt{qvals}, whereas the FWER procedures return
adjusted $p$-values under \texttt{pvals\_adj}. The
Storey--Tibshirani procedure additionally returns the estimated null
proportion under \texttt{pi0}. The Boolean vector \texttt{rejected}
indicates which null hypotheses are rejected at the specified level; a
value of \texttt{False} indicates failure to reject rather than
acceptance of the null hypothesis.

\begin{CodeChunk}
\begin{CodeInput}
>>> import numpy as np
>>> from covtest.multiplicity import fdr, fwer
>>> pvals = np.array([
...     0.001, 0.01, 0.03, 0.20, 0.35,
...     0.45, 0.65, 0.75, 0.85, 0.95,
... ])
>>> weights = np.array([
...     2.0, 2.0, 1.0, 1.0, 1.0,
...     1.0, 1.0, 1.0, 1.0, 1.0,
... ])
>>> res_bh = fdr.benjamini_hochberg(pvals, alpha=0.05)
>>> res_by = fdr.benjamini_yekutieli(pvals, alpha=0.05)
>>> res_wbh = fdr.weighted_bh(pvals, weights, alpha=0.05)
>>> res_storey = fdr.storey_qvalues(pvals, alpha=0.05)
>>> res_holm = fwer.holm(pvals, alpha=0.05)
>>> np.flatnonzero(res_bh["rejected"])
array([0, 1])
>>> np.flatnonzero(res_by["rejected"])
array([0])
>>> np.flatnonzero(res_wbh["rejected"])
array([0, 1])
>>> np.flatnonzero(res_storey["rejected"])
array([0, 1])
>>> round(float(res_storey["pi0"]), 3)
0.737
>>> np.round(res_holm["pvals_adj"][:3], 3)
array([0.01, 0.09, 0.24])
\end{CodeInput}
\end{CodeChunk}

\subsection{Synthetic Data Generation}

\subsubsection{Covariance Matrix Generation}

We provide functions for generating structured covariance matrices, together with a Gaussian Orthogonal Ensemble generator for random-matrix and spectral experiments. The function \code{sample_goe(p)} samples a symmetric random matrix from the Gaussian Orthogonal Ensemble, with off-diagonal entries having variance one and diagonal entries having variance two. Unlike the covariance generators, a GOE matrix is generally not positive definite and is intended for random-matrix and spectral experiments rather than direct use as a covariance matrix. The function  \code{sample\allowbreak\_cov\allowbreak\_uniform\allowbreak\_correlation(K, level)} creates unit-variance correlation matrices with off-diagonal entries drawn uniformly from a range determined by \code{level}. The function \code{generate\allowbreak\_spectral\allowbreak\_cov(p)} constructs covariance matrices with randomly sampled eigenvalues and orthogonal bases, and \code{generate_toeplitz_cov(p, rho)} creates AR(1)-structured Toeplitz matrices with exponentially decaying correlations. For modular dependencies, \code{generate\allowbreak\_block\allowbreak\_diagonal\allowbreak\_cov(p, block_size)} assembles block-diagonal matrices with positive semi-definite blocks, and \code{generate\allowbreak\_low\allowbreak\_rank\allowbreak\_cov(p, rank, noise_var)} combines low-rank structure with isotropic Gaussian noise. To simulate sparse dependencies, \code{generate\allowbreak\_sparse\allowbreak\_precision\allowbreak\_cov(p, sparsity)} samples a sparse precision matrix and returns its inverse covariance. The function \code{generate\allowbreak\_marchenko\allowbreak\_pastur(p, n)} samples from the white Wishart ensemble to approximate the Marcenko–Pastur distribution, while \code{generate\allowbreak\_spiked\allowbreak\_covariance(p, spike\allowbreak\_eigenvalue, num\allowbreak\_spikes)} creates spiked covariance matrices with a low-rank perturbation in the leading eigenspace. Together, these generators provide a flexible testbed for evaluating statistical methods under a range of spectral and structural covariance regimes.

\begin{CodeChunk}
\begin{CodeInput}
>>> import numpy as np
>>> rng = np.random.default_rng(42)
>>> from covtest.simulation import generate_covariances as gc
>>> method  = gc.generate_spiked_covariance
>>> options = dict(spike_eigenvalue=8.0, num_spikes=2)
>>> Sigma   = method(p=5, rng=rng, **options)
\end{CodeInput}
\end{CodeChunk}

\subsubsection{Data Generation}

We provide flexible generators for simulating multivariate heavy-tailed data under both null and alternative covariance structures, potentially with covariance matrices generated as above. The \code{generate\allowbreak\_heavy\allowbreak\_tailed\allowbreak\_samples} function produces samples with exact covariance matching a user-supplied matrix, supporting a wide range of heavy-tailed distributions including Student's t, Laplace, log-normal, Pareto, and scale mixtures. For hypothesis testing or robustness evaluation, \code{generate\allowbreak\_heavy\allowbreak\_tailed\allowbreak\_alternative} introduces structured deviations from the null through mean-shift mixtures, scaled covariance inflation, rank-one perturbations, or spiked eigenvalue bumps, while preserving heavy-tailed behavior. Both functions accept a shared interface with user-defined distribution parameters via options. As an example, 

\begin{CodeChunk}
\begin{CodeInput}
>>> import numpy as np
>>> rng = np.random.default_rng(42)
>>> from covtest.simulation import generate_data as gd
>>> from covtest.simulation import generate_covariances as gc
>>> Sigma = gc.generate_spectral_cov(p=5, rng=rng)
>>> method  = gd.generate_heavy_tailed_samples
>>> options = dict(df=5)
>>> X = method(Sigma, n=100, dist_type="t", rng=rng, options=options)
>>> X = method(Sigma, n=100, dist_type="laplace", rng=rng)
>>> X = method(Sigma, n=100, dist_type="pareto", rng=rng, options={"b": 2.5})
>>> method   = gd.generate_heavy_tailed_alternative
>>> alt_opts = dict(
...     alt="mixture",
...     dist={"df": 5},
...     mu_norm=1.0,
...     p_mix=0.5
... )
>>> result = method(Sigma, n=100, dist_type="t", rng=rng, options=alt_opts)
>>> result["alt"], result["X"].shape, np.round(result["cov_used"], 2)
\end{CodeInput}
\begin{CodeOutput}
('mixture', (100, 5), array([[..., ..., ...], ...]))
\end{CodeOutput}
\end{CodeChunk}

\subsection{Test Diagnostics}

The function \code{analyze_pvalues(pvals, num_permutations, seed)} computes a comprehensive suite of diagnostic statistics to assess whether a vector of $p$-values is consistent with a uniform distribution, as expected under a valid null. It includes classical tests such as the Kolmogorov–Smirnov \citep{smirnov1948table} test and an Anderson–Darling-like statistic tailored to the Uniform[0,1] distribution, along with scalar deviation measures (L$^\infty$ and L$^2$ norms) between the empirical and theoretical CDFs. Additional diagnostics include the genomic inflation factor ($\lambda_\mathrm{GC}$), Storey’s estimator of the null proportion ($\pi_0$), binomial enrichment tests for small-$p$ tails, and a linear regression fit to the QQ plot (slope and intercept). A permutation-based $p$-value is also computed to assess the significance of the observed L$^2$ ECDF deviation. The function returns all results as a structured dictionary, enabling further tabulation or visualization.

\begin{CodeChunk}
\begin{CodeInput}
>>> from covtest.diagnostics.evaluate_pvalues import analyze_pvalues
>>> import numpy as np
>>> pvals_null = np.random.default_rng(0).uniform(0, 1, size=500)
>>> result = analyze_pvalues(pvals_null, num_permutations=1000, seed=123)
>>> result.keys()
\end{CodeInput}
\begin{CodeOutput}
dict_keys(['ks', 'ad', 'ecdf_deviation', 'inflation_factor',
           'storey_pi0', 'tail_tests', 'qq_fit', 'perm_l2_pval'])
\end{CodeOutput}
\end{CodeChunk}
\begin{CodeChunk}
\begin{CodeInput}
>>> from covtest.plotting.null import plot_pvalue_diagnostics_grid
>>> plot_pvalue_diagnostics_grid(pvals_null, sname='Figure3.pdf')
\end{CodeInput}
\end{CodeChunk}

\begin{figure}[t!]
\centering
\includegraphics[width=\textwidth]{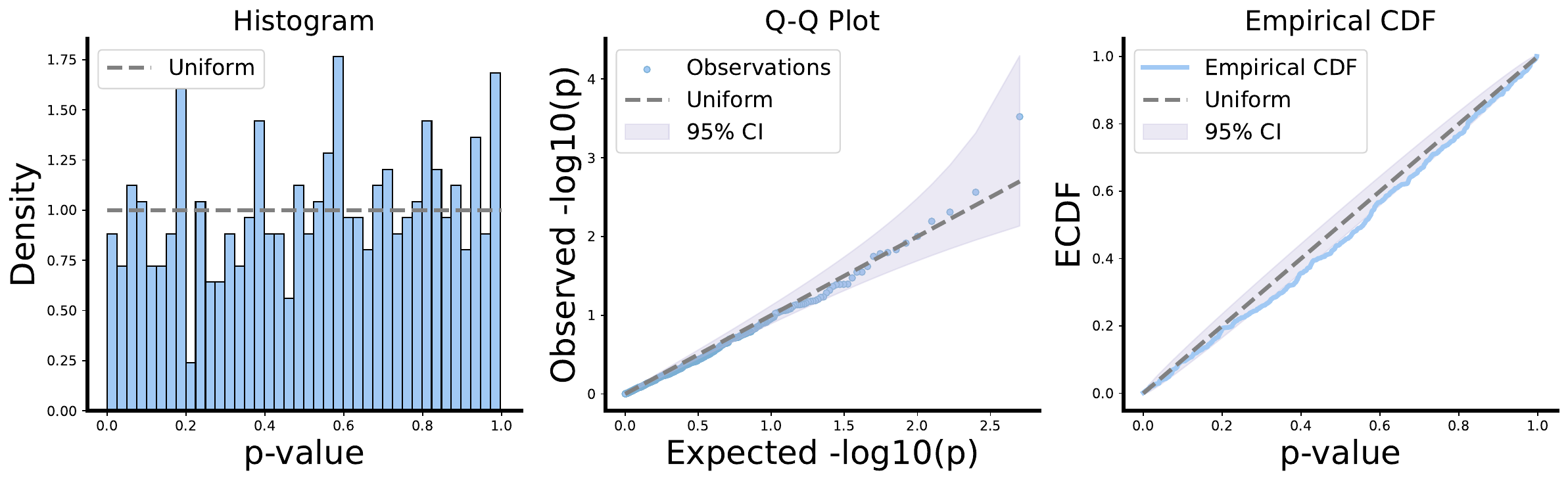}
\caption{Diagnostic plots of the observed $p$-value distribution under the uniform null hypothesis. 
\textbf{Left:} Histogram of the observed $p$-values, with the uniform density shown by the dashed line. 
\textbf{Middle:} Q--Q plot of the observed versus theoretical $-\log_{10}(p)$ quantiles, with the pointwise 95\% confidence band under the uniform null shown in gray. 
\textbf{Right:} Empirical cumulative distribution function (ECDF) of the observed $p$-values, together with the expected uniform CDF and its pointwise 95\% confidence band.}
\label{fig:null_plot}
\end{figure}

An example of the output is given in Figure~\ref{fig:null_plot}. Each of the plots can be reproduced individually as well.

\subsection{Data Loading}

The functions \code{load_iris()}, \code{load_mnist()}, and \code{load_tcga()}
provide convenient access to benchmark datasets stored as compressed NumPy
archives. Only the small Iris archive is bundled inside the wheel; the two
large archives, \file{mnist.npz} and \file{tcga.npz}, are hosted externally
(Zenodo, DOI \code{10.5281/zenodo.21600332}) and are downloaded on first use
via \pkg{pooch}, then cached in the user's operating-system cache directory so
that subsequent calls are immediate. Because \pkg{pooch} is an optional
dependency, the remote datasets require installing the \code{datasets} extra:

\begin{CodeChunk}
\begin{CodeInput}
pip install scikit-covtest[datasets]
\end{CodeInput}
\end{CodeChunk}

Calling \code{load_mnist()} or \code{load_tcga()} without \pkg{pooch}
installed raises an \code{ImportError} with installation instructions.

The function
\code{load_mnist(split="train", return_X_y=True, normalize=True)}
loads the MNIST handwritten digits dataset, returning either the training or
the test split as a pair of arrays, or, when
\code{return_X_y=False}, a dictionary containing both splits. When
\code{normalize=True}, pixel values are cast to single precision and scaled to
the range $[0, 1]$, and images are flattened to vectors of length 784 if they
are stored as $28\times28$ arrays.

The function
\code{load_tcga(return_X_y=True, return_names=False)}
loads the gene-expression dataset described in
Section~\ref{sec:tcga-example}. By default it returns the feature matrix and
the vector of cancer-type labels. Setting
\code{return_names=True} additionally returns the feature names, the name of
the label column, and the sample identifiers; setting \code{return_X_y=False}
returns the same five arrays as a dictionary.

\begin{CodeChunk}
\begin{CodeInput}
>>> from covtest.datasets import load_mnist, load_tcga
>>> X, y = load_mnist(split="train", normalize=True)
>>> X, y, gene_names, label_names, sample_ids = load_tcga(return_names=True)
\end{CodeInput}
\end{CodeChunk}

As distributed, the first column of the array returned by \code{load_tcga()}
is the sample identifier rather than a gene, so the returned matrix has
$20{,}532$ columns of which $20{,}531$ are gene-expression features. This
column must be removed before numerical analysis; Section~\ref{sec:tcga-example}
shows the required step.

%

\section{Null Calibration and Empirical Power}
\label{sec:simulations}

We evaluate a representative subset of the implemented tests across all
four hypothesis classes using controlled Gaussian simulations.
For each class we assess two properties: (i)~ {size control} under
the null hypothesis, examined via probability--probability (P--P) plots
of empirical against theoretical uniform $p$-values; and
(ii)~ {empirical power} against a fixed alternative as the sample
size increases.
All data are generated using the \pkg{scikit-covtest} simulation
utilities \code{generate\_toeplitz\_cov} and
\code{generate\_spiked\_covariance}.
The nominal level is $\alpha = 0.05$ throughout, and all experiments use
$p = 20$ fixed dimensions and Gaussian data.

\begin{figure}[t]
  \centering
  \includegraphics[width=\textwidth]{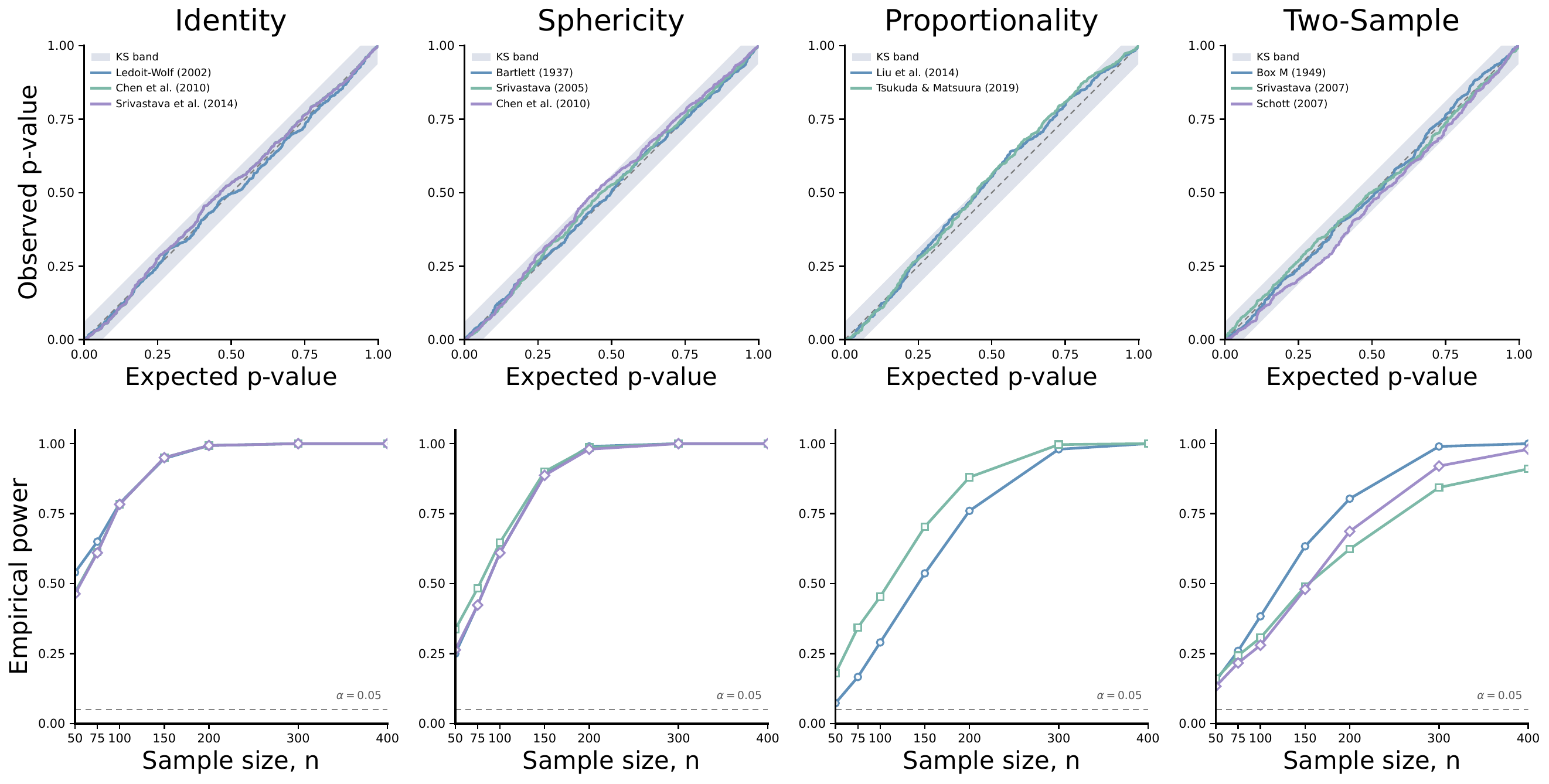}
  \caption{Null calibration  {(top row)} and empirical power
     {(bottom row)} for a representative subset of tests across the
    four hypothesis classes in \pkg{scikit-covtest}, evaluated under
    Gaussian data with $p = 20$ dimensions.
     {Top row}: P--P plots of observed against expected $p$-values
    under the null.
    The shaded region is a 95\% simultaneous Kolmogorov--Smirnov
    confidence band; the dashed line is the $y = x$ reference.
     {Bottom row}: empirical power at
    $\alpha = 0.05$.}
  \label{fig:simulations}
\end{figure}

\subsection{Experimental design}
Null calibration is assessed at $n = 200$ over 500 independent
replications.
Power is estimated over 300 replications at sample sizes
$n \in \{50, 75, 100, 150, 200, 300, 400\}$.
The null and alternative distributions for each class are as follows.

\begin{itemize}
  \item  {Identity.}
    Null: $\mathbf{x} \sim \mathcal{N}(\mathbf{0}, \mathbf{I}_p)$.
    Alternative: $\mathbf{x} \sim \mathcal{N}(\mathbf{0},
    \boldsymbol{\Sigma}_{\mathrm{spike}})$, a spiked model with two
    leading eigenvalues equal to $1.8$ and the remaining $p - 2$
    eigenvalues equal to $1.0$.

  \item  {Sphericity.}
    Null: $\mathbf{x} \sim \mathcal{N}(\mathbf{0}, 3\mathbf{I}_p)$.
    Alternative: $\mathbf{x} \sim \mathcal{N}(\mathbf{0},
    \boldsymbol{\Sigma}_{\mathrm{T}}(0.15))$, an AR(1) Toeplitz matrix
    with off-diagonal decay parameter $\rho = 0.15$, which is not
    proportional to the identity.

  \item  {Proportionality.}
    Null: $(\mathbf{x}, \mathbf{y})$ drawn from
    $\mathcal{N}(\mathbf{0}, \boldsymbol{\Sigma}_{\mathrm{T}}(0.3))$ and
    $\mathcal{N}(\mathbf{0}, 2\boldsymbol{\Sigma}_{\mathrm{T}}(0.3))$
    respectively, satisfying $\boldsymbol{\Sigma}_1 = c\boldsymbol{\Sigma}_2$
    with $c = 1/2$.
    Alternative: $\mathbf{y}$ is redrawn from
    $\mathcal{N}(\mathbf{0}, \boldsymbol{\Sigma}_{\mathrm{T}}(0.12))$,
    which is not proportional to $\boldsymbol{\Sigma}_{\mathrm{T}}(0.3)$
    because the lag-$k$ entry ratio $(0.3/0.12)^k$ varies with $k$.

  \item  {Two-sample.}
    Null: $\boldsymbol{\Sigma}_1 = \boldsymbol{\Sigma}_2 =
    \boldsymbol{\Sigma}_{\mathrm{T}}(0.3)$.
    Alternative: $\boldsymbol{\Sigma}_2 =
    \boldsymbol{\Sigma}_{\mathrm{T}}(0.45)$, a denser correlation
    structure.
\end{itemize}

\subsection{Null calibration}
\label{sec:null-calibration}

Results are shown in the top row of Figure~\ref{fig:simulations}. For the
identity tests, all three methods, \cite{ledoit2002some}, \cite{chen2010tests}, and \cite{srivastava2014tests}, track the diagonal
closely and remain within the 95\% simultaneous Kolmogorov--Smirnov
confidence band, indicating well-controlled Type~I error at this dimension
and sample size. The three sphericity tests \cite{bartlett1937properties}, \cite{srivastava2005some}, and \cite{chen2010tests} are likewise close to the diagonal, with only a
marginal excursion of \citet{chen2010tests} trace at the upper end
of the band.

Among the proportionality tests, both methods are conservative at this
combination of $n$ and $p$: observed $p$-values systematically exceed their
expected uniform quantiles, so both reject less often than the nominal level
warrants. The departure is somewhat more pronounced for
\citet{tsukuda2019high} than for \citet{liu2014new}, though neither procedure
tracks the diagonal as tightly as the identity or sphericity tests in this
regime.

For the two-sample class, \citet{box1953non} is the best calibrated of the
three procedures examined, tracking the diagonal closely across the full range
with no notable excursion outside the confidence band; the $F$ approximation
to the same statistic behaves similarly. \citet{schott2007test} is the only
method whose trace leaves the band, drifting mildly anti-conservative, 
consistent with its reliance on a standard normal limit that is not yet fully
accurate at this combination of dimension and sample size. \citet{srivastava2007multivariate}
trends in the opposite direction, staying within the band but rejecting
somewhat less often than the nominal rate would suggest.

\subsection{Empirical power}
\label{sec:empirical-power}

Power curves are shown in the bottom row of Figure~\ref{fig:simulations}. For
the identity class, the two high-dimensional U-statistic tests \cite{chen2010tests}; \cite{srivastava2014tests} achieve moderately
higher power than Ledoit--Wolf (2002) at small $n$, with all three converging
to near-perfect detection by $n \approx 200$. For sphericity, all three
methods produce nearly indistinguishable power curves; the mild Toeplitz
alternative ($\rho = 0.15$) constitutes a low-magnitude departure from
sphericity that requires $n \approx 300$ for reliable detection regardless of
method.

For proportionality, Tsukuda and Matsuura (2019) attains higher power than
Liu \textit{et al.} (2014) across all sample sizes examined. We report this
difference descriptively: both procedures are conservative under the null at
this $(n, p)$, so the gap cannot be attributed to a calibration asymmetry
between them. For the two-sample class, \cite{box1953non} attains the highest power of the three
procedures from $n = 100$ onward, reaching $0.803$ at $n = 200$ and $1.000$
at $n = 400$, compared with $0.623$ and $0.910$ for \cite{srivastava2007multivariate} and
$0.687$ and $0.980$ for \cite{schott2007test}. Because Box's $M$ holds its nominal
size in this design (Section~\ref{sec:null-calibration}), this advantage
reflects genuine sensitivity rather than size inflation.

\section[Worked Example]
{Worked Example with Included Datasets}
\label{sec:applied}

\subsection{Size and Power with MNIST}

\begin{figure}[t!]
\centering
\includegraphics[width=\textwidth]{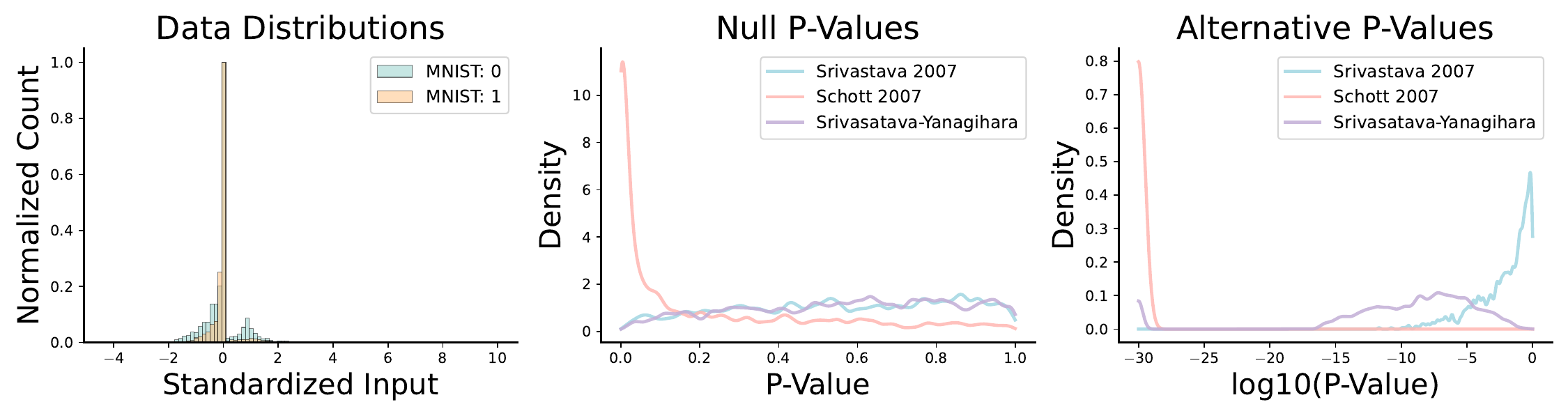}
\caption{\label{fig:mnist} MNIST Two-Sample Test Results. The left plot shows the distribution of the data, while the middle plot shows the distribution of p-values under the null hypothesis for three potential tests. The right panel shows the distribution of log p-values under the alternative, distinguishing the covariances of digits 0 and 1.}
\end{figure}

In our first example, we evaluate covariance testing on the MNIST dataset~\citep{lecun2002gradient}. These data are $28\times 28$ images of digits with 784 values indicating intensity. When vectorized, they yield 784 covariates, making this dataset clearly in the modern high-dimensional regime. We standardize the covariates to have 0 mean and unit variance, then truncate values greater than 10 to 10. The distribution of intensities for 0s and 1s is plotted on the left of Figure~\ref{fig:mnist}. As such, this dataset is ideal to evaluate how non-normality affects two-sample tests on ``real'' data. Code for understanding the critical parts of the simulation is shown below, the full notebook can be found on the GitHub page.

To assess calibration under the null, each replication draws $2N$
distinct digit-0 images without replacement and divides them into two
nonoverlapping groups of size $N$. This construction prevents the same
image from appearing in both samples within a replication. The sample
size $N$ and number of replications are supplied explicitly to the
evaluation function.

\begin{CodeChunk}
\begin{CodeInput}
>>> import numpy as np
>>> from sklearn.preprocessing import StandardScaler
>>> from covtest.datasets.loader import load_mnist
>>> from covtest.methods.hypothesis_two_sample import (
...     srivastava_two_sample_2007,
...     schott2007,
...     srivastava_yanagihara_two_sample,
... )
>>> X_tr, y_tr = load_mnist(split="train")
>>> X_te, y_te = load_mnist(split="test")
>>> X = np.concatenate([X_tr, X_te], axis=0)
>>> y = np.concatenate([y_tr, y_te], axis=0)
>>> X0 = X[y == 0]
>>> X0_std = StandardScaler().fit_transform(X0)
>>> X0_std[X0_std > 10] = 10
>>> methods = [
...     srivastava_two_sample_2007,
...     schott2007,
...     srivastava_yanagihara_two_sample,
... ]
>>> if 2 * N > X0_std.shape[0]:
...     raise ValueError("N is too large to construct disjoint samples.")
>>> p_values = np.empty((n_rep, len(methods)))
>>> rng = np.random.default_rng(0)
>>> for i in range(n_rep):
...     idx = rng.choice(
...         X0_std.shape[0],
...         size=2 * N,
...         replace=False,
...     )
...     subset0 = X0_std[idx[:N]]
...     subset1 = X0_std[idx[N:]]
...     for j, method in enumerate(methods):
...         result = method(subset0, subset1)
...         p_values[i, j] = result["p_value"]
\end{CodeInput}
\end{CodeChunk}

\subsection{Multiple Hypothesis Testing with TCGA}
\label{sec:tcga-example}

In our second example, we demonstrate how \pkg{scikit-covtest} can combine
two-sample covariance-structure tests with multiple-testing procedures. We use
the gene expression cancer RNA-Seq dataset \citep{fiorini2016gene},
distributed by the UCI Machine Learning Repository, which is a random
extraction from the RNA-Seq (HiSeq) PANCAN dataset assembled by The Cancer
Genome Atlas Pan-Cancer Analysis Project \citep{weinstein2013cancer}. The
extraction comprises $801$ samples and $20{,}531$ gene-level expression
features measured on the Illumina HiSeq platform, with samples stored
row-wise. Features carry anonymized placeholder names of the form
\code{gene_XX} rather than gene symbols, so the individual blocks tested below
are not directly interpretable as named biological units.

The $801$ samples span five tumor types: breast invasive carcinoma (BRCA,
$n = 300$), kidney renal clear cell carcinoma (KIRC, $n = 146$), lung
adenocarcinoma (LUAD, $n = 141$), prostate adenocarcinoma (PRAD, $n = 136$),
and colon adenocarcinoma (COAD, $n = 78$). We compare BRCA, the most
frequently represented tumor type, with LUAD, which supplies a second group of
comparable size drawn from a distinct tissue of origin. Both groups have
$n \gg 5$, so each five-dimensional block below is comfortably
well-conditioned in each group.

Our objective is to test for differences in correlation structure rather than
differences in marginal gene-expression variances. We therefore standardize
each feature separately within each cancer type, so that every feature has
sample mean zero and sample variance one in both groups. After this
transformation, the within-group sample covariance matrices are the
corresponding sample correlation matrices. Applying the two-sample procedure to
these standardized observations therefore provides a plug-in test for
differences between the BRCA and LUAD correlation structures.

To formulate a multiple-testing problem, we partition the $20{,}531$ features
into $\lfloor 20{,}531 / 5 \rfloor = 4{,}106$ nonoverlapping blocks of five
consecutive features and apply the two-sample test of
\citet{srivastava2007multivariate} separately to each block. Because
$4{,}106 \times 5 = 20{,}530$, a single trailing feature is left over and is
discarded. This produces one $p$-value for each of the $4{,}106$ feature
blocks. Note that the array returned by \code{load_tcga()} carries the sample
identifier in its first column, which must be dropped before the expression
values are cast to floating point. The corresponding code is:

\begin{CodeChunk}
\begin{CodeInput}
>>> import numpy as np
>>> from tqdm import trange
>>> from sklearn.preprocessing import StandardScaler
>>> from covtest.datasets import load_tcga
>>> from covtest.methods.hypothesis_two_sample import srivastava_two_sample_2007
>>> X_raw, y  = load_tcga()
>>> X         = X_raw[:, 1:].astype(float)   # drop the sample-ID column
>>> X_brca    = X[y == 'BRCA']
>>> X_luad    = X[y == 'LUAD']
>>> Xb_std    = StandardScaler().fit_transform(X_brca)
>>> Xl_std    = StandardScaler().fit_transform(X_luad)
>>> n_groups  = X.shape[1] // 5              # 20531 // 5 = 4106
>>> pvalues   = np.zeros(n_groups)
>>> for i in trange(n_groups):
...     idx        = 5 * i + np.arange(5)
...     res        = srivastava_two_sample_2007(
...                      Xb_std[:, idx],
...                      Xl_std[:, idx])
...     pvalues[i] = res['p_value']
\end{CodeInput}
\end{CodeChunk}

\begin{figure}[t!]
\centering
\includegraphics[width=\textwidth]{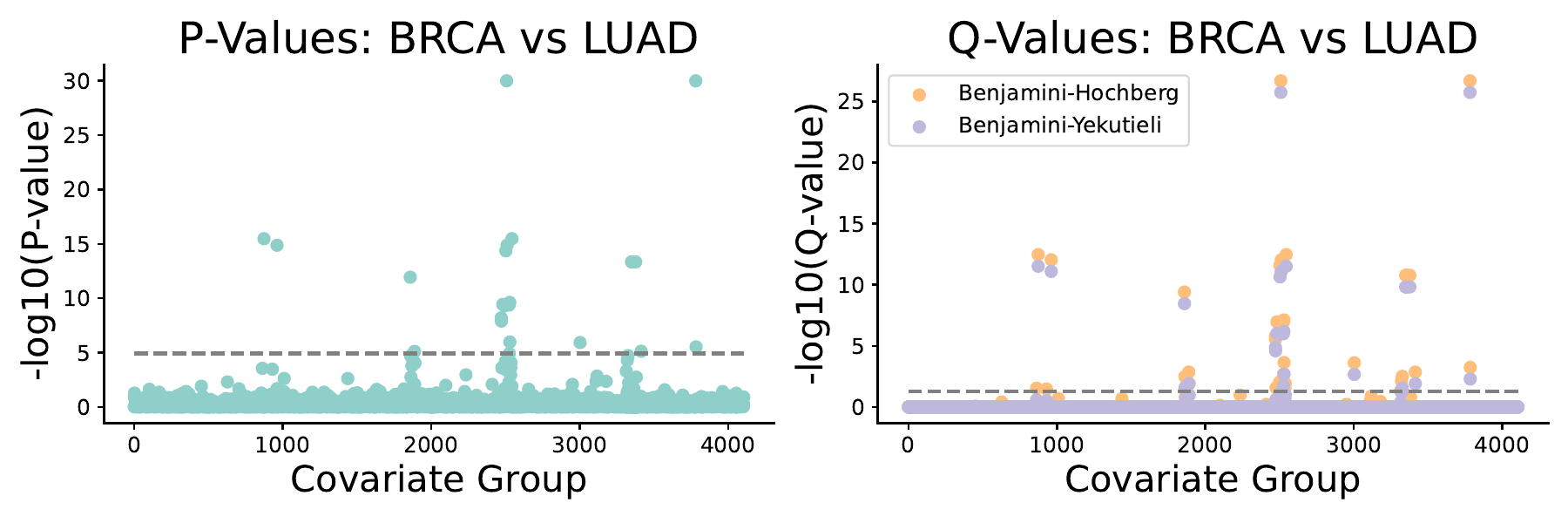}
\caption{\label{fig:tcga} TCGA results. The left panel shows a
Manhattan-style plot of the negative log $p$-values for the $4{,}106$
five-feature blocks in the BRCA-versus-LUAD comparison, with the Bonferroni
threshold for FWER control at the $0.05$ level. The right panel shows the same
$p$-values after adjustment by the two false discovery rate procedures.}
\end{figure}

A plot of the negative log $p$-values is shown on the left of
Figure~\ref{fig:tcga}, together with the Bonferroni threshold for
family-wise error rate (FWER) control at the $0.05$ level, which for
$m = 4{,}106$ blocks is $0.05 / 4{,}106 \approx 1.22 \times 10^{-5}$. Under
this criterion, $23$ feature blocks show a statistically significant difference
in covariance structure. The same $p$-values can be passed directly to
false discovery rate (FDR) procedures, using the methods of
\citet{benjamini1995controlling} and \citet{benjamini2001control}:

\begin{CodeChunk}
\begin{CodeInput}
>>> from covtest.multiplicity.fdr import (
...     benjamini_hochberg,
...     benjamini_yekutieli,
... )
>>> res_bh = benjamini_hochberg(pvalues)
>>> res_by = benjamini_yekutieli(pvalues)
\end{CodeInput}
\end{CodeChunk}

Both adjustments are shown on the right of Figure~\ref{fig:tcga}. The three
procedures control different error criteria, so their rejection counts should
be read as a progression rather than as competing estimates of the same
quantity. Bonferroni controls the FWER, the probability of even one false
rejection, and is the most stringent: it rejects $23$ blocks. The remaining two
procedures control the FDR, the expected proportion of false rejections among
those declared significant, and are correspondingly less stringent. Of these,
\citet{benjamini2001control} is the more conservative, because it deflates the
step-up thresholds by the harmonic factor
$H_m = \sum_{i=1}^{m} 1/i \approx 8.90$ in order to remain valid under
arbitrary dependence among the block-level $p$-values; it rejects $27$ blocks.
The procedure of \citet{benjamini1995controlling}, which assumes independence
or positive regression dependence, applies no such penalty and rejects $42$
blocks. The ordering $23 \le 27 \le 42$ is the expected one: relaxing FWER
control to FDR control, and then dropping the arbitrary-dependence penalty,
can only increase the number of rejections.

\section[Conclusion]{Conclusion}
\label{sec:conclusion}

In this work, we introduce \pkg{scikit-covtest}, a comprehensive \proglang{Python} package for hypothesis testing of covariance matrices. Our primary objective is to provide reliable, well-tested implementations of a wide range of hypothesis tests drawn from the statistical literature. To support researchers and practitioners, we include clear demonstrations using synthetic data, systematic performance comparisons across diverse statistical regimes, and illustrative applications using real datasets.

The package is organized into components to promote clarity and extensibility. The core module, \code{methods}, implements the hypothesis tests. The \code{diagnostics} module provides tools for assessing the plausibility of test assumptions and identifying potential violations. The \code{multiplicity} module addresses multiple hypothesis testing, a critical consideration in high-dimensional and large-scale applications. The \code{simulation} module generates synthetic data under both null models and common alternatives, enabling controlled benchmarking. Finally, the \code{testing} module streamlines systematic evaluations and performance comparisons of selected methods.

Although we incorporated a broad range of available tests, some were undoubtedly omitted. Furthermore, the field is evolving and additional methods will inevitably be developed. We plan to maintain and expand \pkg{scikit-covtest} as new approaches are proposed, and we welcome contributions from the community through our public repository (\url{https://github.com/bystrogenomics/scikit-covtest}).

\bibliography{foo}

\end{document}